\documentclass{PoS}

\usepackage{braket,cases,latexsym,bbold}
\usepackage{color}

\newcommand{\Dcut}{D_{\mathrm{cut}}}
\newcommand{\Tc}{T_{\mathrm{c}}}

\title{Phase transition of four-dimensional Ising model with tensor network scheme}

\ShortTitle{Phase transition of four-dimensional Ising model with TN scheme}

\author{\speaker{Shinichiro Akiyama}\\
        Graduate School of Pure and Applied Sciences, University of Tsukuba, Tsukuba, Ibaraki 305-8571, Japan\\
        E-mail: \email{akiyama@het.ph.tsukuba.ac.jp}}

\author{Yoshinobu Kuramashi\\
        Center for Computational Sciences, University of Tsukuba, Tsukuba, Ibaraki 305-8577, Japan\\
        E-mail: \email{kuramasi@het.ph.tsukuba.ac.jp}}

\author{Takumi Yamashita\\
        Faculty of Engineering, Information and Systems, University of Tsukuba, Tsukuba, Ibaraki 305-8573, Japan\\
        E-mail: \email{yamasita@ccs.tsukuba.ac.jp}}

\author{Yusuke Yoshimura\\
        Center for Computational Sciences, University of Tsukuba, Tsukuba, Ibaraki 305-8577, Japan\\
        E-mail: \email{yoshimur@ccs.tsukuba.ac.jp}}

\abstract{We investigate the phase transition of the four-dimensional Ising model with two types of tensor network scheme, one is the higher-order tensor renormalization group and the other is the anisotropic tensor renormalization group. The results for the internal energy and magnetization obtained by the former algorithm with the impure tensor method, enlarging the lattice volume up to $1024^4$, are consistent with the weak first-order phase transition. For the later algorithm, our implementation successfully reduces the execution time thanks to the parallel computation and the results provided by ATRG seems comparable to those with HOTRG.}

\FullConference{37th International Symposium on Lattice Field Theory - Lattice2019\\
		16-22 June 2019\\
		Wuhan, China}

\begin{document}

\section{Introduction}

According to the perturbative renormalization group analysis, the leading scaling behavior of the four-dimensional ferromagnetic Ising model is specified by the mean-field theory and it is modified by the multiplicative logarithmic factor \cite{PhysRevB.7.248}. Since the Ising model is characterized by the infinite coupling limit of the single-component scalar $\phi^{4}$ theory, the model in four dimensions has been attracting the interest of particle physicists in the context of the triviality of the scalar $\phi^{4}$ theory. Numerical simulation of the Ising model on hypercube lattice therefore serves as a nonperturbative indirect test of the triviality \cite{Kenna:2004cm}, but no Monte Carlo (MC) study has confirmed the logarithmic correction in the scaling behavior of the specific heat, $(\ln|t|)^{1/3}$ with $t$ the reduced temperature. The latest and detailed MC simulation was carried out by Lundow and Markstr\"{o}m and they revealed that the extrapolation to the reliable thermodynamic limits based on the MC study with the linear system size $L\le80$ was hindered by a non-vanishing finite-volume effect \cite{PhysRevE.80.031104,Lundow:2010en}. 

As a different approach other than the MC method, it is so much worth trying the tensor network scheme. In this work we employ the tensor renormalization group approach originally proposed by Levin and Nave~\cite{Levin:2006jai}, which already has wide numerical applications to the field theories in particle physics~\cite{Shimizu:2012zza,Shimizu:2014uva,Shimizu:2014fsa,Takeda:2014vwa,Kawauchi:2016xng,Shimizu:2017onf,Kadoh:2018hqq,Unmuth-Yockey:2018ugm,Kadoh:2018tis,Kuramashi:2018mmi,Butt:2019uul,Kuramashi:2019cgs}, to investigate the four-dimensional Ising model. This is a kind of real-space renormalization group approach, and some of the algorithms are ready to be applied to higher-dimensional systems allowing a direct treatment of the huge lattice, essentially in the thermodynamic limit.
One of such methods is the {\it higher-order tensor renormalization group} (HOTRG) \cite{PhysRevB.86.045139}, which has been applied to many lattice systems, including the three-dimensional Ising model \cite{PhysRevB.86.045139,Wang_2014}. The other is the {\it anisotropic tensor renormalization group} (ATRG) recently proposed by Adachi {\it et~al.} \cite{Adachi:2019paf}. The biggest advantage of ATRG is the drastic reduction of memory and computational costs, compared with HOTRG, and ATRG is potentially able to achieve the higher accuracy than HOTRG with the fixed execution time \cite{Adachi:2019paf}. We apply the  HOTRG and ATRG with parallel computation to investigate the phase transition of the four-dimensional Ising model.


\section{HOTRG and ATRG with parallel computation}

\begin{figure}[th]
\centering\includegraphics*[width=.6\hsize]{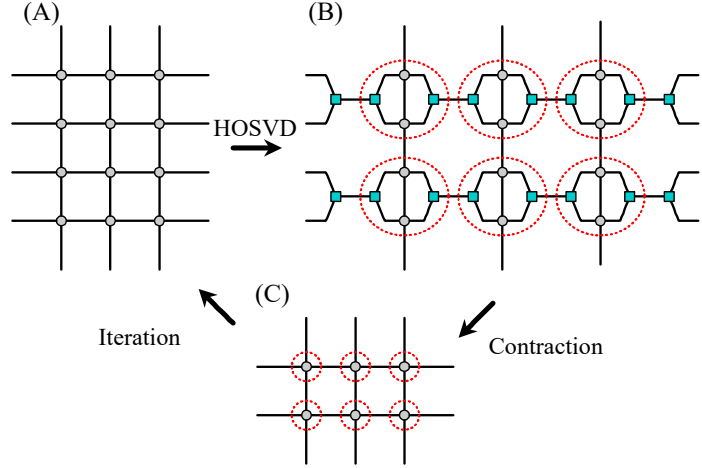}
\caption{Schematic illustration of HOTRG. (A) Local tensors construct tensor network representation. (B) Higher-order SVD (HOSVD) introduces the optimal approximation. (C) After the contraction, the lattice size is reduced by a factor of 2.}
\label{fig:hotrg}
\end{figure}

\begin{figure}[th]
\centering\includegraphics*[width=.6\hsize]{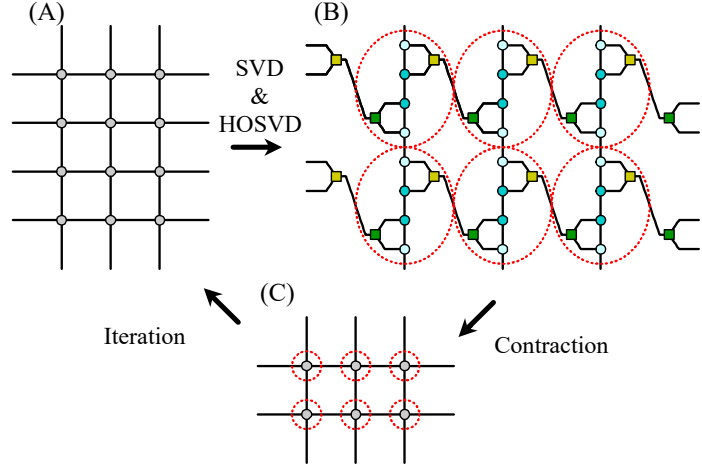}
\caption{Schematic illustration of ATRG. (A) Local tensors construct tensor network representation. (B) SVD for local tensor introduces the extra approximation compared with HOTRG. (C) After the contraction, the lattice size is reduced by a factor of 2.}
\label{fig:atrg}
\end{figure}


The algorithms of HOTRG and ATRG are schematically illustrated in Figs.~\ref{fig:hotrg} and \ref{fig:atrg}, respectively. The approximation applied in HOTRG is based only on the higher-order singular value decomposition (HOSVD) for the two adjacent local tensors, but ATRG has two types of approximation; one is the singular value decomposition (SVD) for each local tensor and the other is HOSVD for two adjacent decomposed tensors, which are constructed from the former approximation. 

We briefly explain the efficiency of the parallel computation for HOTRG. The HOTRG originally proposed in Ref.~\cite{PhysRevB.86.045139} is easily extended to the $d$-dimensional lattice system and its memory cost scales with $\Dcut^{2d}$ and computational time $\Dcut^{4d-1}$, where $\Dcut$ is the bond dimension which controls the accuracy of the HOTRG algorithm. The calculation of $\mathcal{O}(\Dcut^{4d-1})$ is devoted to the tensor contraction to complete the block-spin transformation illustrated with the red dotted circle in Fig~\ref{fig:hotrg} (B), which can be carried out with parallel computation. Indeed, the implementation proposed in Ref.~\cite{Yamashita} reduces the memory cost per process to $\mathcal{O}(\Dcut^{2d-1})$ and the computational time per process to $\mathcal{O}(\Dcut^{4d-3})$. A key idea of this implementation can also be found in Ref.~\cite{Akiyama:2019xzy}.

We now explain our strategy to develop the four-dimensional ATRG. The proposed algorithm in Ref.~\cite{Adachi:2019paf} includes two types of $\mathcal{O}(\Dcut^{2d+1})$ calculation; one is the partial SVD for swapping tensor indices, which gives the tensors represented with dark blue symbols in Fig.~\ref{fig:atrg} (B) and the other is the contraction to complete the block-spin transformation illustrated with the red dotted circle in Fig~\ref{fig:atrg} (B). For the former computation, Ref.~\cite{Oba:2019csk} points out a way to reduce the cost to $\mathcal{O}(\Dcut^{\mathrm{max}(d+3,7)})$. The authors in Ref.~\cite{Adachi:2019paf} also demonstrate a different technique to reduce the cost of partial SVD to $\mathcal{O}(\Dcut^{d+3})$. On the other hand, the execution time of tensor contraction is reduced by parallel computation. We employ $2\Dcut$ processes to distribute the elements of the two locally decomposed tensors represented with light blue symbols in Fig.~\ref{fig:atrg} (B) to each process according to the one of $d+1$ indices, which is not contracted in the block-spin transformation. Therefore, the computational cost per process is reduced from $\mathcal{O}(\Dcut^{2d+1})$ to $\mathcal{O}(\Dcut^{2d})$. The memory footprint of our implementation scales similarly to the original ATRG as $\Dcut^{d+1}$. We have implemented the randomized SVD (RSVD) for the partial SVD. For the detailed algorithm of RSVD, see Ref.~\cite{JSSv089i11}, for instance. The accuracy of RSVD is controlled by the over sampling parameter $p$ and the numbers of the iteration $q$ to update the orthogonal matrix, which is used to reduce the size of the matrix to be decomposed. RSVD can reproduce the comparable result to the ordinal SVD with sufficiently large $p$ and $q$ unless the degenerated singular values are decimated. 

\section{Numerical results}

\subsection{HOTRG}

\begin{figure}[th]
\begin{tabular}{p{0.45\textwidth}p{0.03\textwidth}p{0.45\textwidth}}
\centering\includegraphics*[width=.45\columnwidth,keepaspectratio,clip]{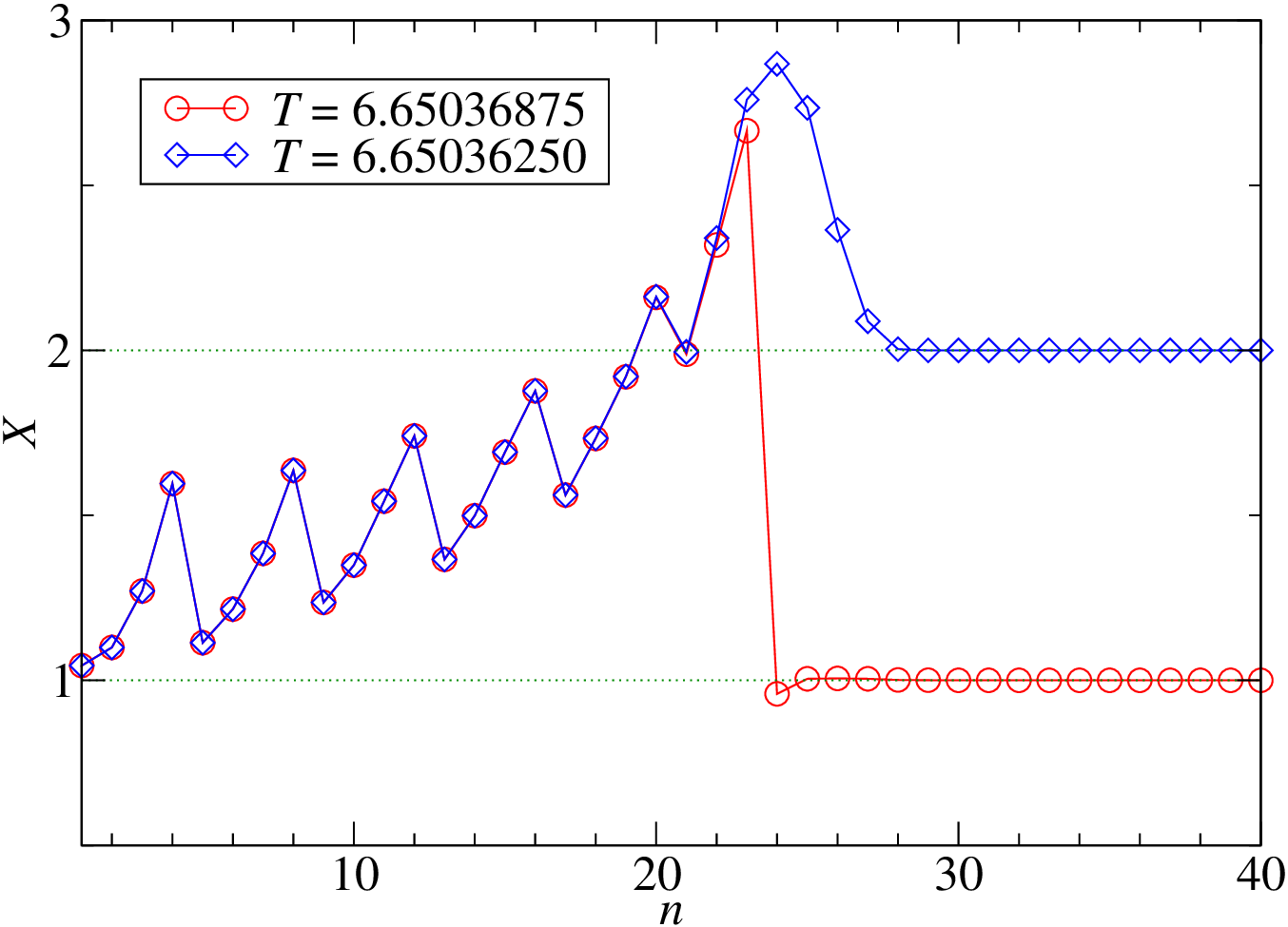}
\caption{$X$ at the $n$-th iteration of HOTRG with $\Dcut=13$. Red and blue lines correspond to the disordered and ordered phases, respectively. }
\label{fig:x}
&&
\centering\includegraphics*[width=.45\columnwidth,keepaspectratio,clip]{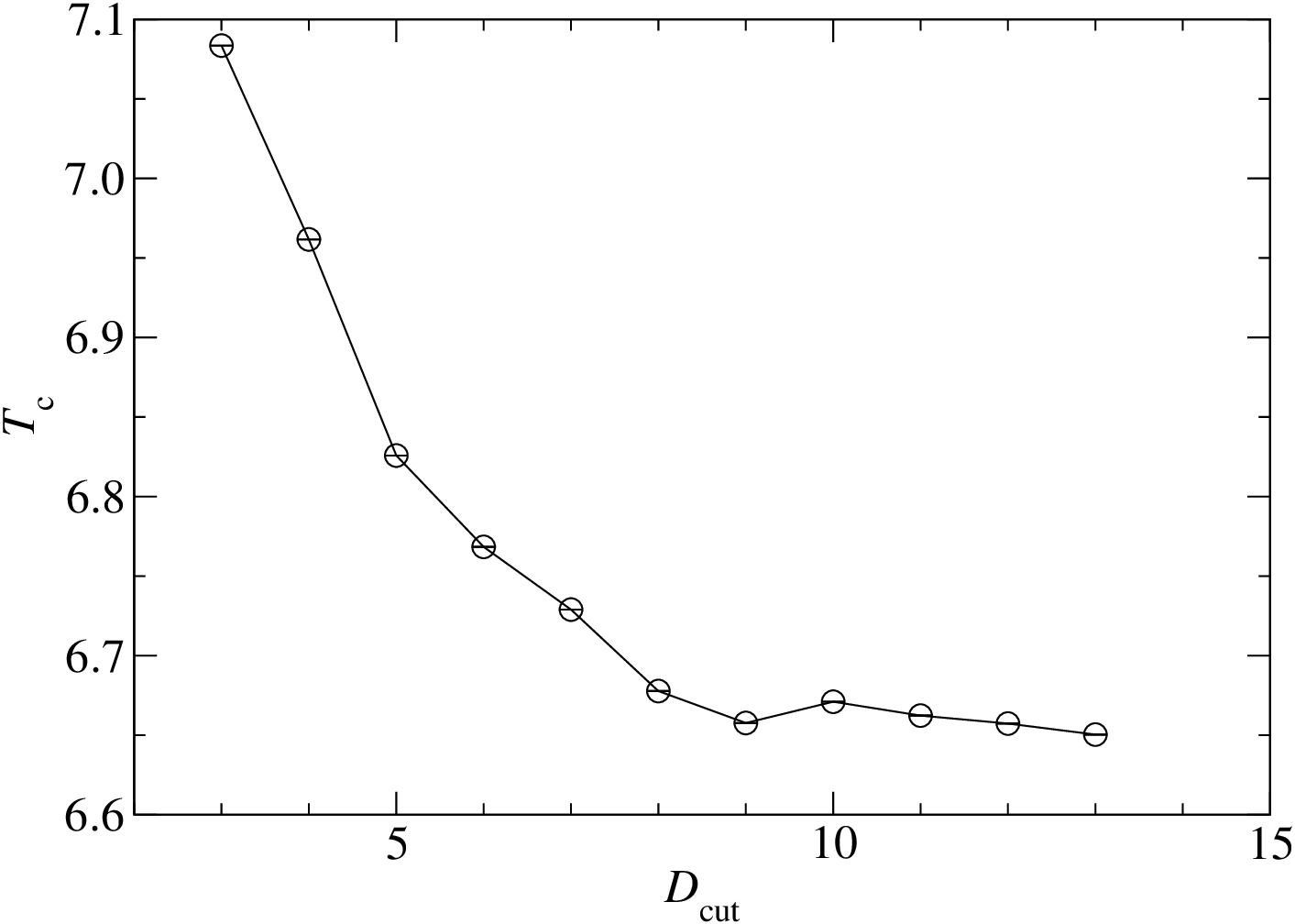}
\caption{Transition point as a function of bond dimension. Error bars are within the symbols.}
\label{fig:tc}
\end{tabular}
\begin{tabular}{p{0.45\textwidth}p{0.03\textwidth}p{0.45\textwidth}}
\centering\includegraphics*[width=.45\columnwidth,keepaspectratio,clip]{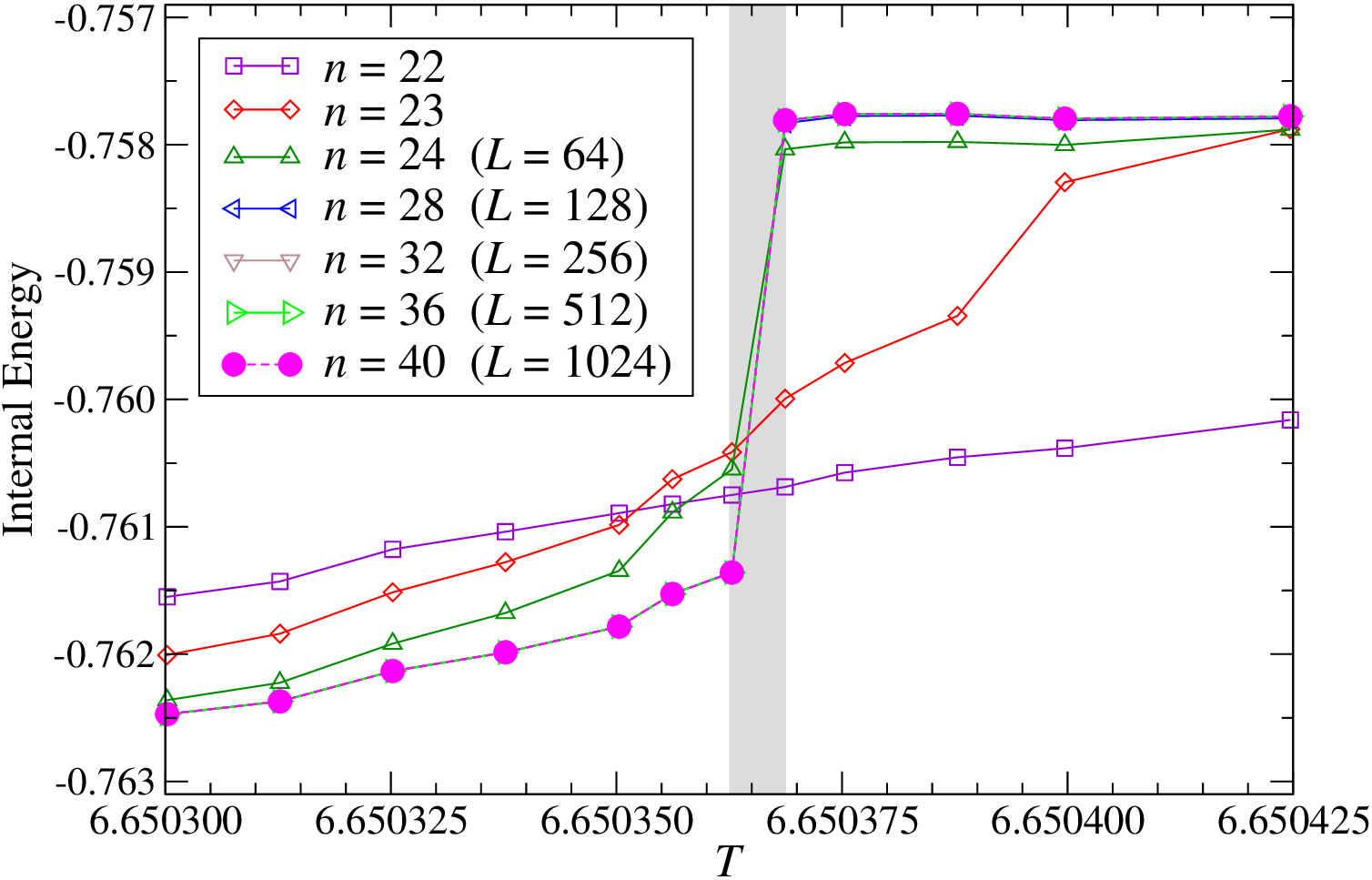}
\caption{Internal energy obtained by HOTRG with $\Dcut=13$. The lattice volume reaches $2^{n}$ after $n$ iterations of HOTRG. $\Tc$ estimated by $X$ is located within the gray band. }
\label{fig:energy}
&&
\centering\includegraphics*[width=.45\columnwidth,keepaspectratio,clip]{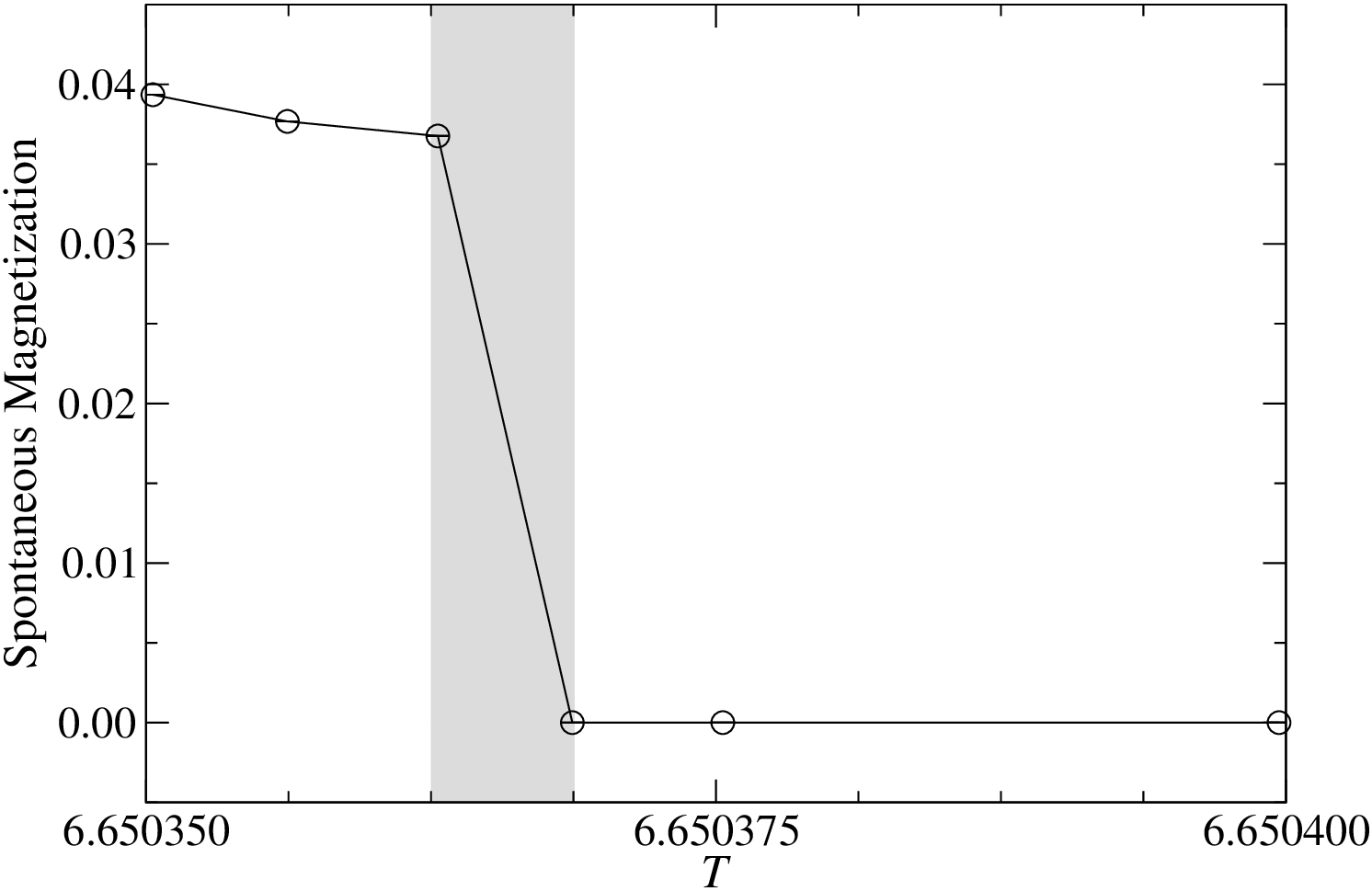}
\caption{Spontaneous magnetization in the thermodynamic limit obtained by HOTRG with $\Dcut=13$. Gray band shows the restriction from $X$ to the location of transition point.}
\label{fig:mag}
\end{tabular}
\end{figure}

We present the results with the parallelized HOTRG, enlarging the lattice size up to $1024^4$. Figure~\ref{fig:x} shows a typical convergence behavior of the indicator of symmetry breaking, referred as $X$ approximately evaluating the degeneracy of the largest components of local tensor, defined in Ref.~\cite{PhysRevB.80.155131}. We can restrict the location of the transition point from the behavior of $X$ (Fig.~\ref{fig:tc}).  As a result, we have obtained $\Tc(\Dcut=13,L\to\infty)=6.650365(5)$. Figure~\ref{fig:energy} shows the internal energy as a function of temperature, where we find a finite jump with mutual crossings of curves for different volumes around the transition point. These are the characteristic features of the first-order phase transition, as discussed in Ref.~\cite{Fukugita1990}. The value of the latent heat in the thermodynamic limit is estimated as $0.0034(5)$. The similar volume dependence of internal energy with a finite jump in the thermodynamic limit has been confirmed with $\Dcut=14$.  We also evaluate the spontaneous magnetization (Fig.~\ref{fig:mag}) and again a finite jump, whose value is $0.037(2)$, emerges around the transition point. The detailed strategy, referred as the impure tensor method, to evaluate these thermodynamic quantities is described in Ref.~\cite{Akiyama:2019xzy}.

\subsection{ATRG}

We firstly applied ATRG to the two-dimensional Ising model and evaluated the indicator of spontaneous symmetry breaking $X$ and the internal energy with the impure tensor method. The results verified that $X$ and the impure tensor method nicely worked within the algorithm of ATRG. We also confirmed that in the four-dimensional Ising model, the RSVD with $p\ge2\Dcut$ and $q\ge\Dcut$ gave $p,q$-independent values for free energy, as claimed in Ref.~\cite{Oba:2019csk}. For the internal energy, $p,q$-dependence almost vanishes with slightly larger $p$ and $q$ compared with the case of free energy. We always draw entries of the trial matrix in RSVD from the normal distribution $\mathcal{N}(0,1)$ and set $p=4\Dcut$ and $q=2\Dcut$ in the following\footnote{Notice that the procedure to update the orthogonal matrix with $q$ iterations can be carried out within $\mathcal{O}(q\Dcut^{\mathrm{min}(d+3,6)})$ computational complexity, employing the technique proposed in Ref.~\cite{Oba:2019csk}.}. We ascertained that the execution time for our implementation of the four-dimensional ATRG scaled with $\Dcut^{8}$.

\begin{figure}[th]
\begin{tabular}{p{0.45\textwidth}p{0.03\textwidth}p{0.45\textwidth}}
\centering\includegraphics*[width=.45\columnwidth,keepaspectratio,clip]{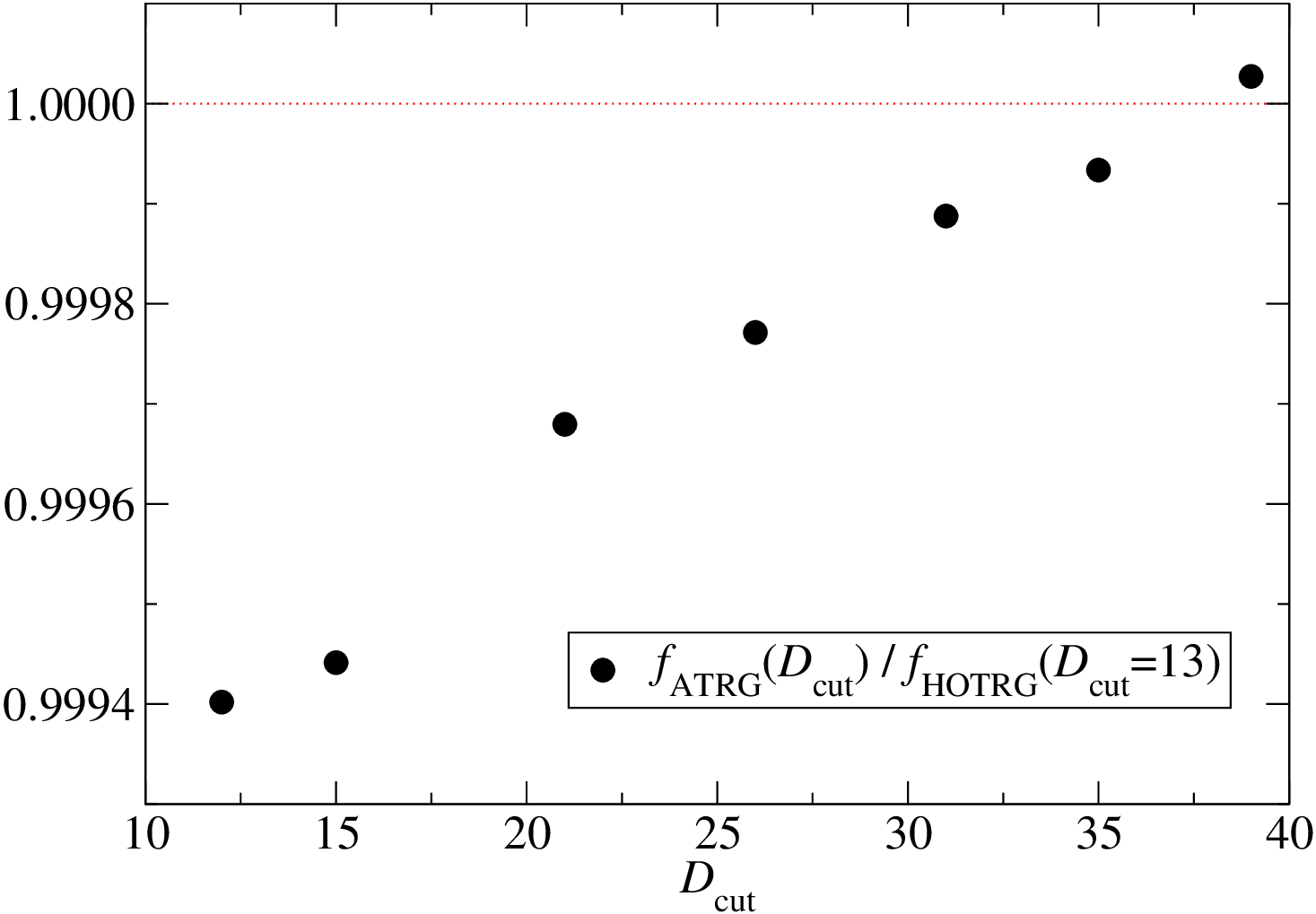}
\caption{Ratio of the free energy at $T=6.65035$ and $L=1024$ obtained by ATRG and HOTRG. Horizontal axis shows the bond dimension set in ATRG.}
\label{fig:atrg1}
&&
\centering\includegraphics*[width=.45\columnwidth,keepaspectratio,clip]{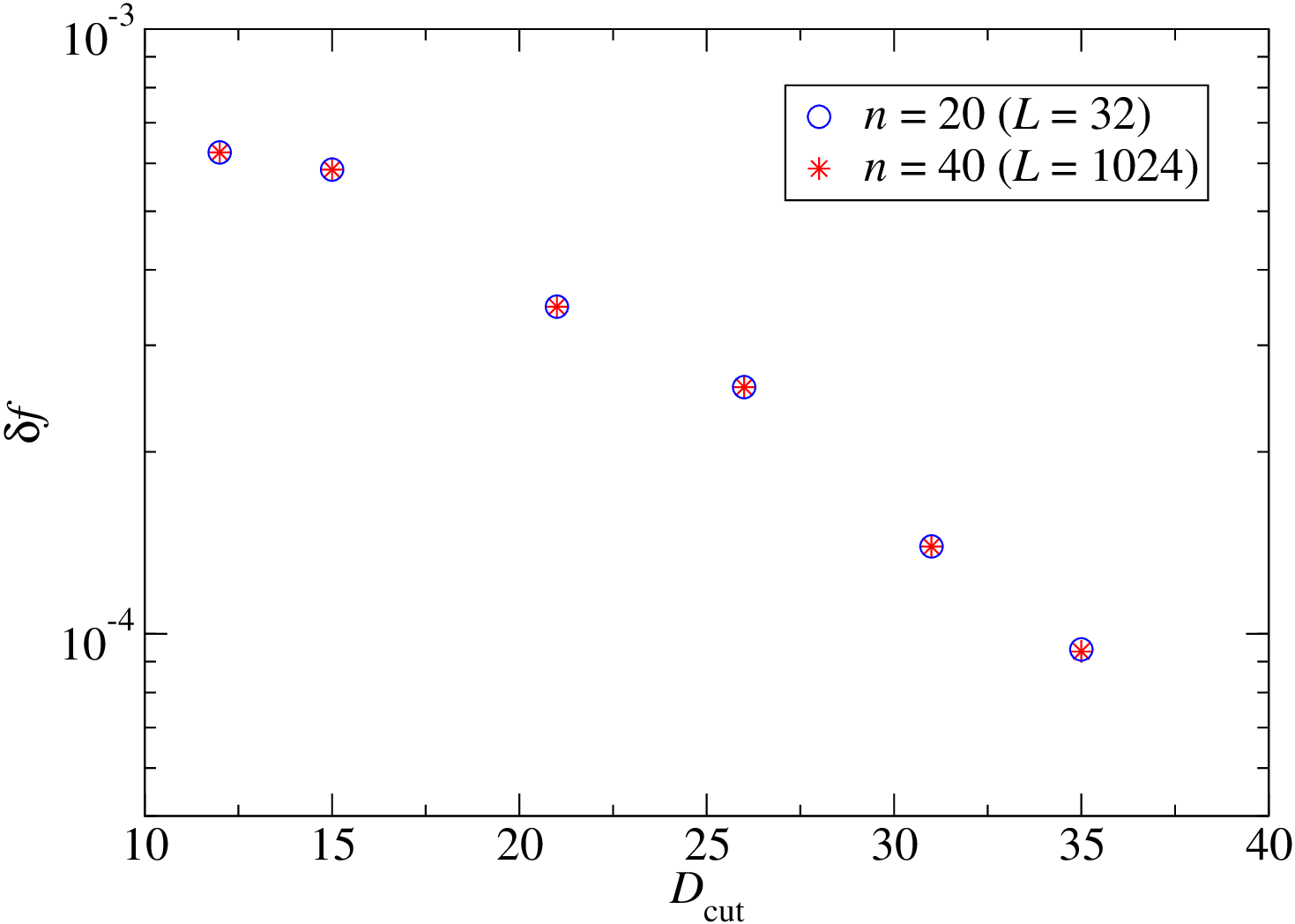}
\caption{Convergence behavior of free energy at $T=6.65035$ as a function of bond dimension.}
\label{fig:atrg2}
\end{tabular}
\begin{tabular}{p{0.45\textwidth}p{0.03\textwidth}p{0.45\textwidth}}
\centering\includegraphics*[width=.45\columnwidth,keepaspectratio,clip]{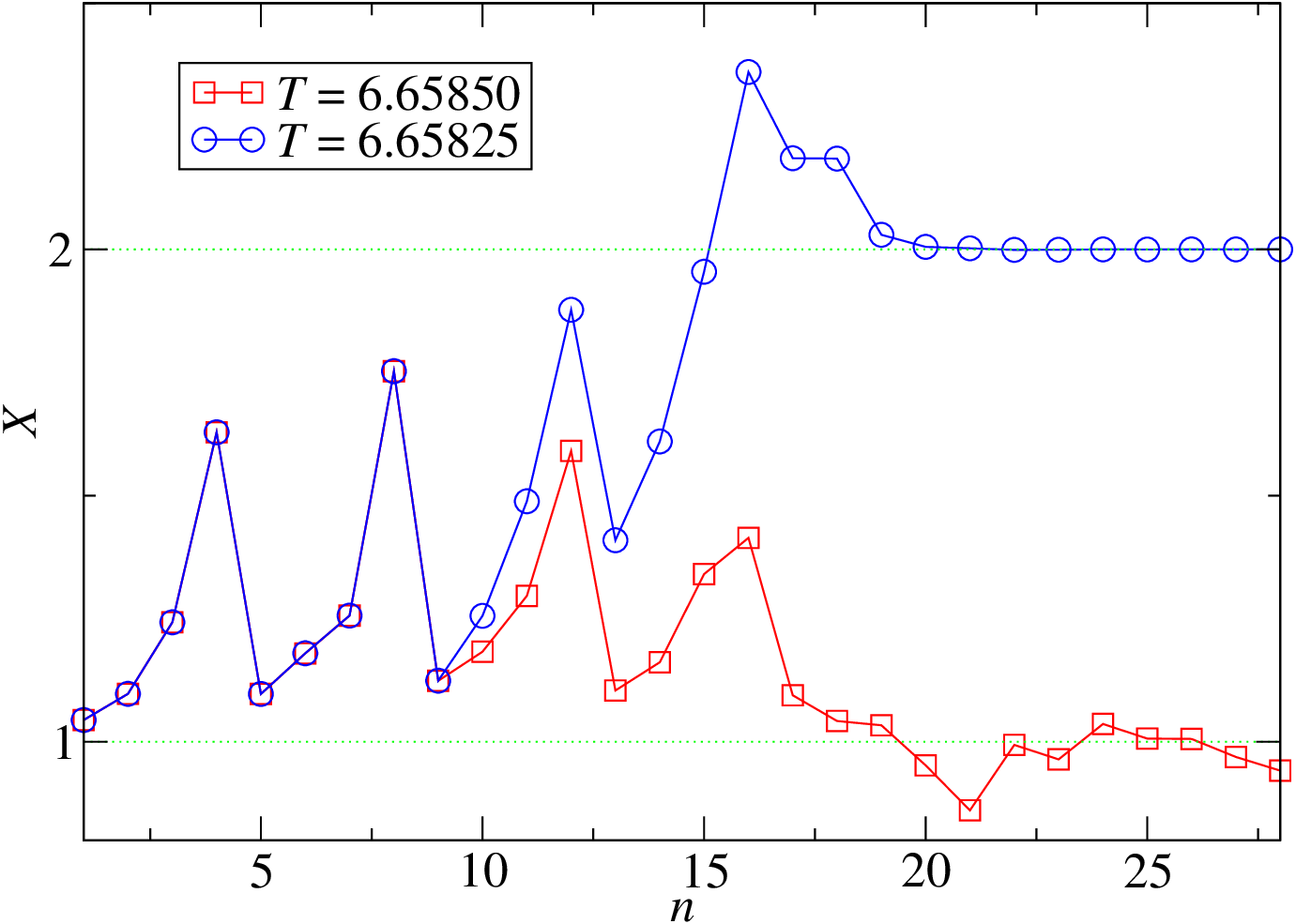}
\caption{$X$ at the $n$-th iteration of ATRG with $\Dcut=39$. Red and blue lines correspond to the disordered and ordered phases, respectively. }
\label{fig:atrg3}
&&
\centering\includegraphics*[width=.45\columnwidth,keepaspectratio,clip]{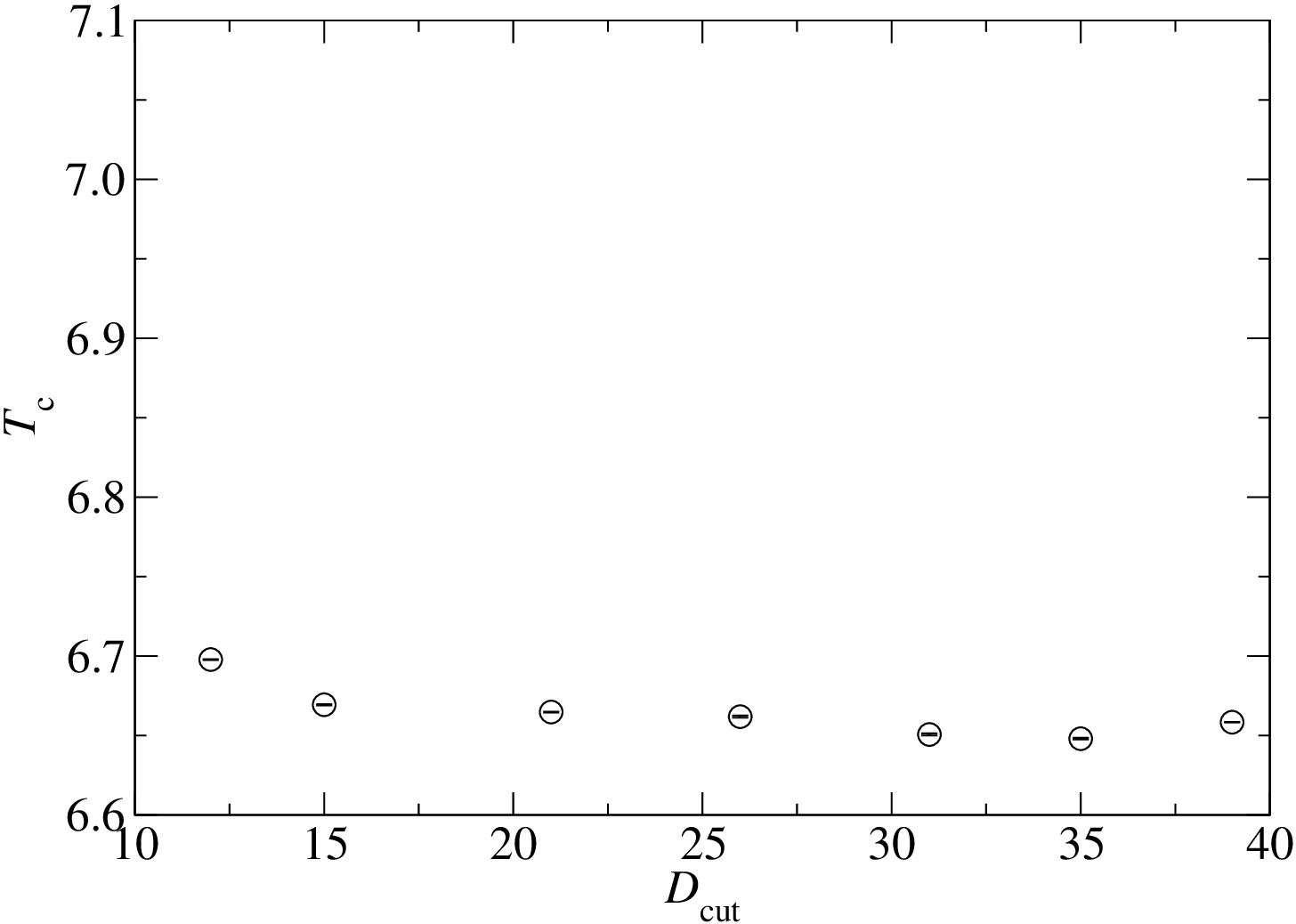}
\caption{Transition point as a function of bond dimension. Error bars are all within the symbols.}
\label{fig:atrg4}
\end{tabular}
\end{figure}

We now move on to the preliminary results for the four-dimensional Ising model obtained by the parallelized ATRG. We have varied the bond dimension in ATRG up to $39$. At the same bond dimension, the accuracy of ATRG should be degraded compared with that of HOTRG, because of the extra approximation introduced in ATRG. For comparison, we observe the free energy normalized by that of HOTRG with $\Dcut=13$; $f_{\mathrm{ATRG}}(\Dcut)/f_{\mathrm{HOTRG}}(\Dcut=13)$. The result at $L=1024$ is shown in Fig.~\ref{fig:atrg1}. We found that our implementation of ATRG with $\Dcut=39$ tended to take almost the same execution time to obtain the comparable value of free energy by our parallelized HOTRG with $\Dcut=13$. Figure~\ref{fig:atrg2} shows
\begin{equation}
\delta f=\left|f_{\mathrm{ATRG}}(\Dcut)-f_{\mathrm{ATRG}}(\Dcut=39)\right|/\left|f_{\mathrm{ATRG}}(\Dcut=39)\right|, 
\end{equation}
which estimates the $\Dcut$ dependence in free energy. Though $\delta f$ decreases monotonically as a function of bond dimension, the convergence of free energy seems slightly slower compared with the $\Dcut=13$ case in HOTRG \cite{Akiyama:2019xzy}. Figure~\ref{fig:atrg3} shows the behavior of $X$ evaluated by ATRG with $\Dcut=39$. 
With the use of $X$, we specify the location of transition temperature as a function of bond dimension (Fig.~\ref{fig:atrg4}). The difference between $\Tc(\Dcut=13)$ by HOTRG and $\Tc(\Dcut=39)$ by ATRG is about $0.12\%$. 
We have also evaluated the internal energy. Although it shows a signal of the finite jump at the transition point as in Fig.~\ref{fig:energy}, it may need further investigation taking account of the current situation that we have not yet confirmed a sufficient convergence of the free energy in terms of $\Dcut$.    



\section{Summary and outlook}

We have studied the phase transition of the four-dimensional Ising model with two types of tensor network scheme, HOTRG and ATRG. The results obtained by HOTRG are consistent with the characteristic features of the weak first-order phase transition; a finite jump in the internal energy with the mutual crossings of curves for different volumes and the discontinuity emerged in the order parameter around the transition point. The ATRG, whose computational cost is so fascinating in application of tensor network scheme to the higher-dimensional systems, has also gave the comparable results to HOTRG for the free energy and the transition point. Further investigation of the internal energy with ATRG is currently in progress.

\clearpage

\section*{Acknowledgement}

Numerical calculation for the present work was carried out with the Oakforest-PACS system of Joint Center for Advanced High Performance Computing under the Interdisciplinary Computational Science Program of Center for Computational Sciences, University of Tsukuba. This work is supported by the Ministry of Education, Culture, Sports, Science and Technology (MEXT) as "Exploratory Challenge on Post-K Computer (Frontiers of Basic Science: Challenging the Limits)".

\bibliographystyle{JHEP}
\bibliography{forthispaper,trg,hotrg,4dising}

\providecommand{\href}[2]{#2}\begingroup\raggedright\begin{thebibliography}{10}

\bibitem{PhysRevB.7.248}
F.~J. Wegner and E.~K. Riedel,
  \href{https://doi.org/10.1103/PhysRevB.7.248}{\emph{Phys. Rev. B} {\bfseries
  7} (1973) 248}.

\bibitem{Kenna:2004cm}
R.~Kenna, \href{https://doi.org/10.1016/j.nuclphysb.2004.05.012}{\emph{Nucl.
  Phys.} {\bfseries B691} (2004) 292}
  [\href{https://arxiv.org/abs/hep-lat/0405023}{{\ttfamily hep-lat/0405023}}].

\bibitem{PhysRevE.80.031104}
P.~H. Lundow and K.~Markstr{\"o}m,
  \href{https://doi.org/10.1103/PhysRevE.80.031104}{\emph{Phys. Rev. E}
  {\bfseries 80} (2009) 031104}.

\bibitem{Lundow:2010en}
P.~H. Lundow and K.~Markstr{\"o}m,
  \href{https://doi.org/10.1016/j.nuclphysb.2010.12.002}{\emph{Nucl. Phys.}
  {\bfseries B845} (2011) 120}
  [\href{https://arxiv.org/abs/1010.5958}{{\ttfamily 1010.5958}}].

\bibitem{Levin:2006jai}
M.~Levin and C.~P. Nave,
  \href{https://doi.org/10.1103/PhysRevLett.99.120601}{\emph{Phys. Rev. Lett.}
  {\bfseries 99} (2007) 120601}
  [\href{https://arxiv.org/abs/cond-mat/0611687}{{\ttfamily
  cond-mat/0611687}}].

\bibitem{Shimizu:2012zza}
Y.~Shimizu, \href{https://doi.org/10.1142/S0217732312500356}{\emph{Mod. Phys.
  Lett.} {\bfseries A27} (2012) 1250035}.

\bibitem{Shimizu:2014uva}
Y.~Shimizu and Y.~Kuramashi,
  \href{https://doi.org/10.1103/PhysRevD.90.014508}{\emph{Phys. Rev.}
  {\bfseries D90} (2014) 014508}
  [\href{https://arxiv.org/abs/1403.0642}{{\ttfamily 1403.0642}}].

\bibitem{Shimizu:2014fsa}
Y.~Shimizu and Y.~Kuramashi,
  \href{https://doi.org/10.1103/PhysRevD.90.074503}{\emph{Phys. Rev.}
  {\bfseries D90} (2014) 074503}
  [\href{https://arxiv.org/abs/1408.0897}{{\ttfamily 1408.0897}}].

\bibitem{Takeda:2014vwa}
S.~Takeda and Y.~Yoshimura,
  \href{https://doi.org/10.1093/ptep/ptv022}{\emph{PTEP} {\bfseries 2015}
  (2015) 043B01} [\href{https://arxiv.org/abs/1412.7855}{{\ttfamily
  1412.7855}}].

\bibitem{Kawauchi:2016xng}
H.~Kawauchi and S.~Takeda,
  \href{https://doi.org/10.1103/PhysRevD.93.114503}{\emph{Phys. Rev.}
  {\bfseries D93} (2016) 114503}
  [\href{https://arxiv.org/abs/1603.09455}{{\ttfamily 1603.09455}}].

\bibitem{Shimizu:2017onf}
Y.~Shimizu and Y.~Kuramashi,
  \href{https://doi.org/10.1103/PhysRevD.97.034502}{\emph{Phys. Rev.}
  {\bfseries D97} (2018) 034502}
  [\href{https://arxiv.org/abs/1712.07808}{{\ttfamily 1712.07808}}].

\bibitem{Kadoh:2018hqq}
D.~Kadoh, Y.~Kuramashi, Y.~Nakamura, R.~Sakai, S.~Takeda and Y.~Yoshimura,
  \href{https://doi.org/10.1007/JHEP03(2018)141}{\emph{JHEP} {\bfseries 03}
  (2018) 141} [\href{https://arxiv.org/abs/1801.04183}{{\ttfamily
  1801.04183}}].

\bibitem{Unmuth-Yockey:2018ugm}
J.~Unmuth-Yockey, J.~Zhang, A.~Bazavov, Y.~Meurice and S.-W. Tsai,
  \href{https://doi.org/10.1103/PhysRevD.98.094511}{\emph{Phys. Rev.}
  {\bfseries D98} (2018) 094511}
  [\href{https://arxiv.org/abs/1807.09186}{{\ttfamily 1807.09186}}].

\bibitem{Kadoh:2018tis}
D.~Kadoh, Y.~Kuramashi, Y.~Nakamura, R.~Sakai, S.~Takeda and Y.~Yoshimura,
  \href{https://doi.org/10.1007/JHEP05(2019)184}{\emph{JHEP} {\bfseries 05}
  (2019) 184} [\href{https://arxiv.org/abs/1811.12376}{{\ttfamily
  1811.12376}}].

\bibitem{Kuramashi:2018mmi}
Y.~Kuramashi and Y.~Yoshimura,
  \href{https://doi.org/10.1007/JHEP08(2019)023}{\emph{JHEP} {\bfseries 08}
  (2019) 023} [\href{https://arxiv.org/abs/1808.08025}{{\ttfamily
  1808.08025}}].

\bibitem{Butt:2019uul}
N.~Butt, S.~Catterall, Y.~Meurice and J.~Unmuth-Yockey,
  \href{https://arxiv.org/abs/1911.01285}{{\ttfamily 1911.01285}}.

\bibitem{Kuramashi:2019cgs}
Y.~Kuramashi and Y.~Yoshimura,
  \href{https://arxiv.org/abs/1911.06480}{{\ttfamily 1911.06480}}.

\bibitem{PhysRevB.86.045139}
Z.~Y. Xie, J.~Chen, M.~P. Qin, J.~W. Zhu, L.~P. Yang and T.~Xiang,
  \href{https://doi.org/10.1103/PhysRevB.86.045139}{\emph{Phys. Rev. B}
  {\bfseries 86} (2012) 045139}.

\bibitem{Wang_2014}
S.~Wang, Z.-Y. Xie, J.~Chen, B.~Normand and T.~Xiang,
  \href{https://doi.org/10.1088/0256-307x/31/7/070503}{\emph{Chinese Physics
  Letters} {\bfseries 31} (2014) 070503}.

\bibitem{Adachi:2019paf}
D.~Adachi, T.~Okubo and S.~Todo,
  \href{https://arxiv.org/abs/1906.02007}{{\ttfamily 1906.02007}}.

\bibitem{Yamashita}
T.~Yamashita and T.~Sakurai, {\emph{{in preparation}} }.

\bibitem{Akiyama:2019xzy}
S.~Akiyama, Y.~Kuramashi, T.~Yamashita and Y.~Yoshimura,
  \href{https://doi.org/10.1103/PhysRevD.100.054510}{\emph{Phys. Rev.}
  {\bfseries D100} (2019) 054510}
  [\href{https://arxiv.org/abs/1906.06060}{{\ttfamily 1906.06060}}].

\bibitem{Oba:2019csk}
H.~Oba,  \href{https://arxiv.org/abs/1908.07295}{{\ttfamily 1908.07295}}.

\bibitem{JSSv089i11}
N.~Erichson, S.~Voronin, S.~Brunton and J.~Kutz,
  \href{https://doi.org/10.18637/jss.v089.i11}{\emph{Journal of Statistical
  Software, Articles} {\bfseries 89} (2019) 1}.

\bibitem{PhysRevB.80.155131}
Z.-C. Gu and X.-G. Wen,
  \href{https://doi.org/10.1103/PhysRevB.80.155131}{\emph{Phys. Rev. B}
  {\bfseries 80} (2009) 155131}.

\bibitem{Fukugita1990}
M.~Fukugita, H.~Mino, M.~Okawa and A.~Ukawa,
  \href{https://doi.org/10.1007/BF01334757}{\emph{Journal of Statistical
  Physics} {\bfseries 59} (1990) 1397}.

\end{thebibliography}\endgroup

\end{document}